\documentstyle[12pt]{article}

\catcode`\@=11
\long\def\@makefntext#1{
\protect\noindent \hbox to 3.2pt {\hskip-.9pt
$^{{\ninerm\@thefnmark}}$\hfil}#1\hfill}                

\def\@makefnmark{\hbox to 0pt{$^{\@thefnmark}$\hss}}  

\def\ps@myheadings{\let\@mkboth\@gobbletwo
\def\@oddhead{\hbox{}
\rightmark\hfil\ninerm\thepage}
\def\@oddfoot{}\def\@evenhead{\ninerm\thepage\hfil
\leftmark\hbox{}}\def\@evenfoot{}
\def\sectionmark##1{}\def\subsectionmark##1{}}

\renewcommand{\thefootnote}{\fnsymbol{footnote}}

\def\sectionc{\@startsection {section}{1}{\z@}{-3.5ex plus -1ex minus 
    -.2ex}{2.3ex plus .2ex}{\bf }}
\def\subsectionc{\@startsection{subsection}{2}{\z@}{-3.25ex plus -1ex minus 
   -.2ex}{1.5ex plus .2ex}{\it }}
\renewcommand{\section}[1]{\sectionc{#1}\hspace*{\parindent}}
\renewcommand{\subsection}[1]{\subsectionc{#1}\hspace*{\parindent}}
\newcounter{appendixc}
\newcounter{subappendixc}[appendixc]
\newcounter{subsubappendixc}[subappendixc]

\renewcommand{\appendix}[1] {\vspace*{0.6cm}
        \refstepcounter{appendixc}
        \setcounter{figure}{0}
        \setcounter{table}{0}
        \setcounter{equation}{0}
        \renewcommand{\thefigure}{\Alph{appendixc}.\arabic{figure}}
        \renewcommand{\thetable}{\Alph{appendixc}.\arabic{table}}
        \renewcommand{\theappendixc}{\Alph{appendixc}}
        \renewcommand{\theequation}{\Alph{appendixc}.\arabic{equation}}
        \noindent{\bf Appendix \theappendixc #1}\par\vspace*{0.4cm}}

\def\abstracts#1{{
        \centering{\begin{minipage}{13.2truecm}
        \footnotesize\baselineskip=13pt\noindent
        \parindent=0pt #1
        \end{minipage}}\par}}

\renewenvironment{thebibliography}[1]
        {\begin{list}{\arabic{enumi}.}
        {\usecounter{enumi}\setlength{\parsep}{0pt}
\setlength{\leftmargin 0.75cm}{\rightmargin 0pt}
         \setlength{\itemsep}{0pt} \settowidth
        {\labelwidth}{#1.}\sloppy}}{\end{list}}

\newcounter{itemlistc}
\newcounter{romanlistc}
\newcounter{alphlistc}
\newcounter{arabiclistc}

\newcommand{\fcaption}[1]{
        \refstepcounter{figure}
        \setbox\@tempboxa = \hbox{\footnotesize Figure~\thefigure. #1}
        \ifdim \wd\@tempboxa > 6in
           {\begin{center}
        \parbox{6in}{\footnotesize\baselineskip=13pt Figure~\thefigure. #1}
            \end{center}}
        \else
             {\begin{center}
             {\footnotesize Figure~\thefigure. #1}
              \end{center}}
        \fi}

\newcommand{\tcaption}[1]{
        \refstepcounter{table}
        \setbox\@tempboxa = \hbox{\footnotesize Table~\thetable. #1}
        \ifdim \wd\@tempboxa > 6in
           {\begin{center}
        \parbox{6in}{\footnotesize\baselineskip=13pt Table~\thetable. #1}
            \end{center}}
        \else
             {\begin{center}
             {\footnotesize Table~\thetable. #1}
              \end{center}}
        \fi}

\def\@citex[#1]#2{\if@filesw\immediate\write\@auxout
        {\string\citation{#2}}\fi
\def\@citea{}\@cite{\@for\@citeb:=#2\do
        {\@citea\def\@citea{,}\@ifundefined
        {b@\@citeb}{{\bf ?}\@warning
        {Citation `\@citeb' on page \thepage \space undefined}}
        {\csname b@\@citeb\endcsname}}}{#1}}

\newif\if@cghi
\def\cite{\@cghitrue\@ifnextchar [{\@tempswatrue
        \@citex}{\@tempswafalse\@citex[]}}
\def\citelow{\@cghifalse\@ifnextchar [{\@tempswatrue
        \@citex}{\@tempswafalse\@citex[]}}
\def\@cite#1#2{{$\null^{#1}$\if@tempswa\typeout
        {IJCGA warning: optional citation argument
        ignored: `#2'} \fi}}

 1
 1
 1

\font\ninerm=cmr9

\begin{document}
\hfill{\small {\bf MKPH-T-96-22}}\\
\begin{center}
\normalsize\bf In--Medium Pion Properties
from Chiral Perturbation Theory\footnote{Talk given at Int.\ Symp.\
Non-Nucleonic Degrees of Freedom Detected in Nucleus, Sept.\ 2-5, 1996
(Osaka, Japan)}
\end{center}
\vspace*{0.6cm}
\centerline{\footnotesize M. KIRCHBACH$^1$ and A. WIRZBA$^2$}
\baselineskip=13pt
\centerline{\footnotesize\it $^1$Institut 
f\"ur Kernphysik, J. Gutenberg Universit\"at, D--55099 Mainz, Germany}
\centerline{\footnotesize\it $^2$Institut f\"ur Kernphysik, TH Darmstadt, 
D-64289 Darmstadt, Germany}

\vspace*{0.6cm}
\abstracts{
\noindent
Two--point functions related to the pion weak decay constant $f_\pi$
are calculated from the generating functional of chiral perturbation
theory in the mean field approximation and the heavy--baryon limit. The
aim is to demonstrate that Lorentz invariance is violated in the
presence of background matter. This fact manifests itself in the
splitting of both $f_\pi$ and the pion mass into uncorrelated time--
and space--like parts.  We emphasize the different in--medium
renormalizations of the correlation functions, show the inequivalence
between the in--medium values of $f_\pi$ deduced from Walecka--type
models, on the one hand, and QCD sum rules, on the other hand, and
elaborate on the importance for some nuclear physics observables.}

\vspace*{0.6cm}
\normalsize\baselineskip=15pt
\setcounter{footnote}{0}
\renewcommand{\thefootnote}{\alph{footnote}}

\noindent
The pion weak decay constant is introduced through the matrix element
of the weak axial current $J_{\mu , 5 }^a = \bar q \gamma_\mu \gamma_5
{\tau^a\over 2} q$ between the pion and the hadronic vacuum state,
$\langle 0|\bar q \gamma_\mu\gamma_5 {\tau^a\over 2} q|\pi^a\rangle =
i f_\pi k_\mu $ , where $q$ denotes the isospin quark doublet.  It can
also be introduced via the axial--current--pseudoscalar (AP)
two--point function $T^{AP}_\mu $ (where $P^a \equiv \bar q i\gamma_5
\tau^a q$ is the pseudoscalar density) according to
\begin{eqnarray}
 T^{AP}_\mu  &=&  \frac{ i}{3} 
G_\pi \int {\rm d}^4x\, e^{ik\cdot x}
 \langle 0|TJ_{\mu ,5}^a(x)\pi^a (0)|0\rangle = 
{ {- i G_\pi f_\pi k_\mu}\over {k^2 } }\, ,
 \\
 if_\pi=\lim_{\stackrel{\vec k=0 }{k_0\to 0}}
 \frac{-k^\mu T^{AP}_\mu}{ G_\pi}
 &=&  
  \frac{1}{3}\int {\rm d}^4 x \,
\langle 0|\lbrack J^a_{0, 5}(x), \pi^a (0)\rbrack |0\rangle
 =  \langle 0|\lbrack Q^a_5(0), \pi^a(0)\rbrack|0\rangle \, ,
\label{AP}
\end{eqnarray}
where $Q_5^a(0)$ is the Fourier transform of the axial charge density
$J_{0, 5}^a (x)$, $G_\pi \equiv \langle 0|\bar q i\gamma_5\tau^a
q|\pi^q \rangle$ is the vacuum coupling constant of the pseudoscalar
density to the pion, and CAC has been exploited.  If one evaluates the
commutator in Eq.\,(\ref{AP}) within the linear $\sigma$ model where
$\lbrack Q_5^a(0),\pi^b(0)\rbrack = -i\sigma \delta^{ab}$, one finds
that $f_\pi = - \langle 0|\sigma|0\rangle$. Obviously, it is the $AP$
correlator that underlies the definition of $f_\pi$ within the linear
$\sigma$ model.  Consider now the axial--current--axial--current (AA)
correlator
\begin{equation} 
T^{AA}_{\mu ,\nu} = {i\over 3} \int {\rm d}^4x\, e^{ik\cdot x} 
\langle 0| TJ^a_{\mu ,5}(x) J^a_{\nu , 5}(0)|0\rangle 
=\lbrace g_{\mu \nu } f^2_\pi +
{{k_\mu k_\nu f^2_\pi}\over {m_\pi^2 -k^2 }}\rbrace \, , 
\end{equation}
and the related Gell-Mann--Oakes--Renner (GOR) correlator\,\cite{KiWi}
\begin{equation}
\lim_{\stackrel{\vec k=0}{k_0\to 0}}
{i\over 3} \int {\rm d}^4x\, e^{ik\cdot x} 
\langle 0| T \partial ^\mu J^a_{\mu ,5}(x) \partial^\nu 
J^a_{\nu , 5}(0)|0\rangle 
= -f_\pi^2 m_\pi^2\, .
\label{AA}
\end{equation}
The l.h.s.\ of Eq.\,(\ref{AA}) can furthermore be evaluated to give
${{(m_u+m_d)}\over 2}\langle 0|\bar u u\mbox{+}\bar d d|0\rangle $.
{}From this, the quantity $f_\pi$ is now extracted as
\begin{equation}
   f_\pi^2 m_\pi^2 = -{{m_u\mbox{+}m_d}\over 2} 
 \langle 0|\bar u u\mbox{+}\bar d d|0\rangle \, 
\equiv - 2m_q \langle 0|\bar q q |0\rangle \, .
\label{GOR_1}
\end{equation}
Obviously, the bare value of $f_\pi$ is independent of the choice of
the two--point function used for its definition.  This is not the case
at finite matter densities, as the various correlation functions have
a different in--medium dependence.  In exploiting the generating
functional, $ e^{i Z_{\rm eff}\lbrack s,p, v_\mu ,a_\mu,\eta , \bar
\eta \,\rbrack}$, of chiral perturbation theory (ChPTh) to order
${\cal O}(Q^2)$ in both the mean field (MF) approximation and the
heavy-baryon formulation we here present the analogous in--medium
renormalized two--point functions.  We start with the chiral effective
lagrangian relevant for the isoscalar low--energy $\pi N$ scattering
(see Refs.\,\cite{GSS,ThWi} for notations, further references and
details)
\begin{eqnarray}
L_{\rm eff}  &=& i\bar N (v\cdot \partial -\sigma_N) N 
+{1\over 2}(\partial _\mu \pi)^2 
-{1\over 2}m_\pi^2\pi^2 \nonumber \\
&& \mbox{}+  {1\over f_\pi^2}
\left \{ {1\over 2}\sigma_N \pi^2 +c_2(v\cdot \partial \pi)^2
 +c_3 (\partial_\mu \pi)^2 \right \}\bar N N
+ j^a\pi^a (1-{{\sigma_N \bar N N}\over {f_\pi^2m_\pi^2}})\, ,
\label{eff_La}
\end{eqnarray}
where $j^a =2Bf_\pi p^a$ and $B=-m_\pi^2 /(m_u\mbox{+}m_d)$.  The
presence of background matter is taken into account by the replacement
of $\bar N N$ in $L_{\rm eff}$ with the matter density $\rho$.  Within
this ChPTh scheme, the calculation of the in--medium two--point
functions is now a well--defined procedure, since Green functions are
obtained by taking functional derivatives of the generating functional
with respect to the sources:
\begin{eqnarray}
T_\mu^{AP} = 
{{\delta^2 Z_{\rm eff}}\over 
{\delta a^a_\mu(-k)\delta p^ a (k)}}|_{a=v=p=0, s={\cal M}}\, ,
&\quad & 
T_{\mu ,\nu}^{AA} =
{{\delta^2 Z_{\rm eff}}\over 
{\delta a^a_\mu(-k)\delta a^ a_\nu (k)}}|_{a=v=p=0, s={\cal M}}
\nonumber\\
T^{PP}& =&
{{\delta^2 Z_{\rm eff}}\over {\delta p^a (-k)\delta p^ a (k)}}|
_{a=v=p=0, s={\cal M}}\, .
\label{func_der}
\end{eqnarray}
The latter correlation function leads to $G_\pi^2 /(m_\pi^2 -k^2)$,
and is thus associated with the pion propagator.
The in--medium $T^{PP}$ is then obtained as\,\cite{ThWi}
\begin{eqnarray}
\frac{i}{3}\int {\rm d}^4x\, 
e^{ik\cdot x}\langle \tilde{0}|T P^a(x) P^a(0)|\tilde{0}\rangle &=&
-{{G_\pi^*\, ^2}\over 
{k_0^2 -\gamma (\rho)\vec{k\, }^2 -m_\pi^*\,^2}}\, ,
\nonumber\\
G_\pi^*\, ^2 =G_\pi^2 
{{(1-{{\sigma_N\rho }\over {f_\pi^2m_\pi^2}})^2}\over 
{1 +{{2(c_2+c_3)\rho}\over f_\pi^2}}} \, , &\quad &
c_2\mbox{+}c_3 = -0.27 \,{\rm fm}\, 
,\quad  c_3 = -0.55\,{\rm fm} \, 
,\nonumber\\
m_\pi^* \, ^2 =
{ {1- {{\sigma_N\rho}\over {f_\pi^2m_\pi^2}} }\over
{1 +{{2(c_2+c_3)\rho }\over f_\pi^2}} }m_\pi^2 \, ,&\quad &
\gamma (\rho ) = {{1 +{{2c_3\rho}\over f_\pi^2}}\over
{1 +{{2(c_2+c_3)\rho}\over f_\pi^2}}}\, , \quad \sigma_N = 0.228\, 
{\rm fm}^{-1}\, .
\label{par_numb}
\end{eqnarray}
The pion field corresponding to the propagator in
Eq.\,(\ref{par_numb}) with vacuum weight for the $k_0^2$ term is the
so--called {\em quasi--pion} field $\widetilde{\phi }^a_\pi$, whereas
the commonly used {\em Migdal field\/}\,\cite{KiRi,KiWi} (denoted by
$\widetilde{\pi}^a$) corresponds to the case when the inverse
propagator has the form $k^2 -m_\pi^2 -\Pi (k_0, \vec{k}\, )$ with
$\Pi (k_0, \vec{k}\, )$ standing for the pion self energy (see
Ref.\,\cite{KiWi} for details).  The relations of the quasi--pion and
the Migdal field to the bare pion field are $\widetilde{\phi }_\pi^a =
{G_\pi \over G_\pi^*} \widetilde{\pi} ^a$ and $\widetilde{\pi}^a = (1-
{{\sigma_N\rho }\over {f_\pi^2m_\pi^2}})\pi^a$, respectively.  Note
that ${m_\pi^*}^2$ from Eq.\,(\ref{par_numb}) corresponds to the pole
with respect to a purely time--like pion momentum and is slightly
enhanced ($\approx 8\%$) at nuclear matter density, $\rho_0$, over its
bare value $m_\pi^2$.  If we had considered a purely space--like pion
momentum there, the mass--pole would be placed at ${m_\pi^*}^2 = -
m_\pi^2 (1\mbox{$-$}\frac{\sigma_N\rho}{f_\pi^2m_\pi^2})/ (
1\mbox{+}\frac{2c_3\rho}{f_\pi^2}) \approx -4m_\pi^2\, .  $ The
absence of a common mass--pole signals a violation of Poincar\'{e}
invariance in the presence of background matter and indicates that the
in--medium hadronic states are no longer eigenstates to $p^\mu p_\mu
$.  Similarly, the considered currents do not any longer transform
according to the $\lbrace {1\over 2}, {1\over 2}\rbrace $
representation of the Lorentz group O(3,1), but split into
uncorrelated charge (O(3) singlet) and space--like current (O(3)
triplet) parts\,\cite{KiRi}.  The in--medium $AP$--correlator
evaluated along the line of ChPTh\,\cite{WiTh,KiWi} reads
\begin{equation}
\frac{i}{3}\int {\rm d}^4 x\, e^{ik\cdot x}
\langle \tilde{0}|TJ_{ 0 ,5}^a (x) P^a (0)|\tilde{0}\rangle 
= G_\pi {{ik^0 f_\pi (1- {{\sigma_N\rho}\over {f_\pi^2m_\pi^2}})}
\over {m_\pi^* \, ^2 -k^2_0 +\gamma (\rho) \vec {k\, }^2 }}\, .
\label{AP_1}
\end{equation}
Therefore the in--medium pion weak decay constant
associated with $T^{AP}$ is now 
\begin{equation}
f_\pi^{\rm AP}(\rho )
=\lim_{{\vec{k}}\to 0;{m_\pi^\ast}^2\to 0}\, \frac{i k_\mu}{3 G_\pi}\,
{{\delta^2 Z_{\rm MF}\over 
{\delta a_\mu^a(-k)\, \delta p^a(k) }}}|_{a_\mu=v_\mu=p=0,s={\cal M}}
= f_\pi\, \left(1 -{{\sigma_N\rho }
\over {f_\pi^2m_\pi^2}}\right )\, ,
\label{fpi_AP}
\end{equation}
which exactly reproduces the in--medium expectation value of the
$\sigma$ field in Walecka--type models.  Finally, the in--medium
$T^{AA}_{00}$ is evaluated as\,\cite{ThWi}
\begin{equation}
T^{AA}_{00}(\rho \not= 0) 
= f_\pi^2 \left(1 + {{2(c_2\mbox{+}c_3)\rho }/ f_\pi^2}\right)
 \left(1+ { k_0^2 \left\{{m_\pi^*\, ^2 -k_0^2 +\gamma (\rho )
\vec {k\, }^2} \right\}^{-1}}
\right) \, .
\label{AA_1}
\end{equation}
{}From that in combination with Eq.\,(\ref{GOR_1}) and the in--medium
version of (\ref{AA}) the in--medium GOR relation follows
as\,\cite{ThWi}
\begin{equation}
{f_\pi^{(t)}}^2 {m_\pi ^*}^2 = -2m_q
\langle \tilde{0}|\bar q q|\tilde{0}\rangle
= f_\pi^2 m_\pi^2 
\left(1\mbox{$-$}{{\sigma_N\rho }\over {f_\pi^2 m_\pi^2}}\right)\, 
 , \quad
{f_\pi^{(t)}}^2 
 = f_\pi^ 2\,  \left(1\mbox{+}{{2(c_2\mbox{+}c_3)\rho }\over f_\pi^2} 
 \right)\, .
\label{med_GOR}
\end{equation}
The in--medium pion weak decay constant $f_\pi^{(t)}$ associated with
the GOR relation is in fact independent of the choice for the
in--medium pion field\,\cite{ThWi}.  The question is now which
in--medium weak pion decay constant enters the various matrix elements
between the pion and the dense vacuum states.  As function of the
choice for the pion (quasi--pion, Migdal's) field, the following
expressions hold at nuclear matter density
\begin{eqnarray}
\langle \tilde{0}|J_{\mu ,5}^a|\tilde{\pi}^a\rangle 
= (if_\pi^S\, k_0, if_\pi^P \, \vec{k}\, ) &, &
f_\pi^S =f_\pi
{ {1 +{ {2(c_2+c_3)\rho }\over f_\pi^2}}\over
{1- { {\sigma_N\rho}\over {f_\pi^2m_\pi^2}}}}   \approx 0.9f_\pi\, ,\\ 
f_\pi^P\! =\!f_\pi
{{1{+}{{2c_3\rho}\over f_\pi^2}}\over 
{1{-}{ {\sigma_N\rho}\over {f_\pi^2m_\pi^2} }}}
\!\approx\! {1\over 4}f_\pi  &,&
\langle \tilde{0}|J^a_0 |\tilde{\phi }_\pi^a \rangle \!=\!
f_\pi\sqrt{1\mbox{+}{{2(c_2\mbox{+}c_3)\rho}\over f_\pi^2}} ik_0
\equiv f_\pi^{(t)}ik_0\, .
\label{mael}
\end{eqnarray}
These equations show that one has to insist on the correct combination
of the weak pion decay constant and the pion propagator in
calculating, say, the renormalization of the amplitude
$g_p\mbox{=}f_\pi g_{\pi NN}/ (k^2\mbox{$-$}m_\pi^2 ) $ of the induced
pseudoscalar part of the weak axial nucleon current.{}For example, for
a process of an almost pure space--like momentum transfer (like
ordinary muon capture) it is $f_\pi^P$ which should be combined with
Migdal's propagator. In contrast to that for the case of an almost
pure time--like momentum transfer (like radiative muon capture) the
allowed combinations are either $f_\pi^S$ with Migdal's propagator or
$f_\pi^{(t)}$ with the quasi--pion propagator.  Finally, we would like
to evaluate the density dependence of the amplitude of the two--body
$(2b)$ axial charge operator (see\,\cite{KiRiTsu} for notations)
\begin{equation}
J_{0,5}^a (2b) = C_{\rm MEC}^{\rho =0}
                   \sum_{i\not=j}(\vec{\sigma}_i 
 +\vec{\sigma}_j)\cdot \hat{r}_{ij}
                    Y_1(m_\pi r)(\vec{\tau_i}
 \times \vec{\tau_j})^a\, ,\quad
 C_{\rm MEC}^{\rho =0 } =  
{ {g_{\pi NN}^2\, m_\pi^2} \over {8\pi g_A m_N^2}}\, .
\label{MEC}
\end{equation}
The only way to consistently evaluate the $\rho$ dependence of $J_{0,
5}^a (2b)$ is to express from the beginning the various couplings
entering $C_{\rm MEC}^{\rho }$ by means of QCD sum
rules. Manipulations in $C_{\rm MEC}^{\rho =0}$ by means of the
Goldberger--Treiman (GT) relation, with the aim first to work
$f_\pi^2$ into the denominator and only then consider its $\rho$
dependence, lead to a spurious enhancement of the 2b--axial charge,
because the GT relation is not compatible with the functional
dependence of $g_{\pi NN}$ on the quark condensate, $\langle 0|\bar q
q|0\rangle = -1.48 \,{\rm fm}^{-3}$.  The dependence of $g_{\pi NN}$,
$m_N $, and $g_A$ on $\langle 0|\bar q q|0\rangle $ follows from QCD
sum rules as\,\cite{Reinders}
\[
{g_{\pi NN}^2\over {4\pi }} =2^5\pi^3 {{f_\pi^2 }/ m_N^2}\, ,
\quad 
m_N^3 =-8\pi^2 \langle 0|\bar q q|0\rangle \, ,\quad
g_A\mbox{$-$}1 \approx 0.13 \,{\rm fm}^6\, 
\langle 0|\bar q q|0\rangle^2 \, \approx 0.3\, .
\]
Inserting these quantities into Eq.\,(\ref{MEC}) and in subsequently
replacing the bare by the in--medium quark condensate in accordance
with Eq.\,(\ref{med_GOR}), one finds $C_{\rm MEC}^{\rho=\rho_0} /
C_{\rm MEC}^{\rho =0} \approx (\langle 0| \bar q q|0\rangle /
\langle \tilde{0}| \bar q q| \tilde{0}
\rangle ) ^{{1\over 3}} =(1\mbox{$-$}\sigma_N\rho
/m_\pi^2f_\pi^2)^{-{1\over 3}} \approx 1.15 \, $.  This result shows
that the $40\%$ enhancement of $C_{\rm MEC}^\rho$ (needed to explain
the acceleration of isovector first--forbidden weak decays in the lead
region) requires a strong presence of the scalar--isoscalar
2$\pi$--exchange channel in the NN potential as suggested
in\,\cite{KiRiTsu}.  To conclude, we would like to stress once more
that it is crucial to evaluate the in--medium couplings without
violating Ward identities that have their roots in the current quark
dynamics.

\end{document}